\documentclass{elsart}

\newcommand{\be}{\begin{equation}}
\newcommand{\ee}{\end{equation}}
\newcommand{\ben}{\begin{eqnarray}}
\newcommand{\een}{\end{eqnarray}}

\newcommand{\nd}{{\noindent}}
\usepackage{color}

\usepackage{graphicx}
\usepackage{amsmath}

\begin{document}
\begin{frontmatter}

\title{Deformed Tsallis-statistics analysis of a complex nonlinear matter-field system}

\author[fisica,CIC]{A.M. Kowalski\corauthref{cor}},
\ead{kowalski@fisica.unlp.edu.ar} \corauth[cor]{Corresponding
author.}
\author[fisica,CONICET] {A. Plastino},
\ead{plastino@fisica.unlp.edu.ar}
\address[fisica]{Instituto de F\'isica (IFLP-CCT-Conicet), Fac. de Ciencias Exactas,\\
Universidad Nacional de La Plata,  C.C. 727, 1900 La Plata,
Argentina}
\address[CIC]{Comision de Investigaciones Cient\'ificas (CICPBA)}
\address[CONICET]{Argentina's National Research Council (CONICET)}

\begin{abstract}
\vskip 3mm We study, using information quantifiers,  the dynamics generated by a special Hamiltonian that gives a detailed account of  the interaction between  a classical and a quantum system. The associated, very rich dynamics displays periodicity, quasi-periodicity, not-boundedness, and chaotic regimes. Chaoticity, together with complex behavior, emerge in the proximity of an unstable entirely quantum instance. Our goal is to compare the statistical description provided by Tsallis quantifiers vis a vis  that obtained with  Shannon's entropy and Jensen's complexity.

\vskip 3mm
\nd Keywords: Tsallis Entropy, q-Statistics, Complexity, Semiquantum dynamics, Bandt-Pompe's  probabilities extraction.

\end{abstract}
\end{frontmatter}

\vspace{-0.5cm}\section{Introduction}

\nd Quantifiers derived from information theory, like entropic forms
and statistical complexities (see as examples
\cite{Shannon48,Shiner99,LMC95,Lamberti04}) have been seen  to be very useful for understanding the dynamics connected to time
series, following the  work of  Kolmogorov and Sinai,
who transformed Shannon's information theory into a powerful tool for
the analysis of dynamical systems \cite{Kolmogorov58,Sinai59}. Of course,
information theory measures and probability spaces $\Omega$ are
inseparably joined quantifiers. For obtaining information quantifiers (IQ) one needs first of all to
determine the probability distribution $P$ that characterizes  the
dynamical system or time series under scrutiny.
Many techniques have been proposed for the election of $P \in
\Omega$. We can mention approaches based on symbolic dynamics
\cite{Mischaikow99}, Fourier analysis \cite{Powell79}, and the wavelet
transform \cite{Colo2001}, for example. Bandt and Pompe (BP) \cite{Bandt2002,ours} proposed a symbolic formalism for
finding the probability distribution (PD) $ P$ associated to an
arbitrary time series (see Appendix). BP's approach relied on  peculiar traits of
the attractor-construction problem through  causal
information, that BP  include in  building up the  PD on is looking for.
 A notable BP-result is significant
performance-improvement with regards to the IQs
 one finds  by using their PD-determination methodology. One just has  to
 assume 1) stationarity and 2) that a sufficient data-amount is some  available. \vskip 3mm

 \subsection{Deformed $q$-statistics}

\nd It is  a well-known
fact that physical systems that are characterized by either
long-range interactions, long-term memories, or multi-fractal
nature, are best described  by a generalized statistical mechanics'
formalism \cite{Hanel} that was proposed 30 years ago: the so-called
Tsallis' or $q$-statistics. More precisely, Tsallis~\cite{paper:tsallis1988}
advanced in 1988 the idea of using in a thermodynamics' scenario an
entropic form, the Harvda-Chavrat one, characterized by the entropic
index $q \in \mathcal{R}$ ($q =1$ yields the orthodox Shannon
measure):
\begin{equation}
S_q = \frac{1}{(q-1)} \sum _{i=1}^{N_s} \left[p_i - (p_i)^q\right],
\label{eq:t-entropy}
\end{equation}
where $p_i$ are the probabilities associated  with the associated
$N_s$ different system-configurations. The entropic index (or
deformation parameter) $q$ describes the deviations of Tsallis
entropy from the standard Boltzmann-Gibbs-Shannon-one
\begin{equation}
S = - \sum _{i=1}^{N_s} p_i \ln (p_i). \label{Shannon-entropy}
\end{equation}
  It is well-known that the orthodox entropy works best in
dealing with systems composed of either independent subsystems or
interacting via short-range forces whose subsystems can access all
the available phase space \cite{Hanel}. For systems exhibiting
long-range correlations, memory, or fractal properties,  Tsallis'
entropy becomes the most appropriate mathematical
form~\cite{paper:alemany1994,paper:tsallis1995,paper:tsallis1998,paper:kalimeri2008}.

\subsection{Our semi-quantum physics model}

\nd Now, a topic of great interest is that of the interplay between quantum and classical systems, sometimes called semiquantum physics.
 If quantum effects in one of the systems are small vis-a-vis those of the other, regarding it  as classical not only simplifies the description but provides profound insight into the composite system's dynamics.  One may cite as illustrations the  Bloch equations \cite{Bloch}, two-level systems interacting with an electromagnetic field within a cavity,  the Jaynes-Cummings semi-classical model \cite{Milonni,Sa.91}, collective nuclear features \cite{Ring}, etc.

\nd The system studied here \cite{KR.18} is of interest in both Quantum Optics and Condensed Matter \cite{Milonni,Sa.91,K0,K1}, particularly in view of the fact that we deal with a bosonic system that admits quasi-periodic and unbounded  regimes, separated by an unstable region  \cite{RK.05}. This feature makes the interaction with a classical mode a quite attractive phenomenon. This system has already been studied using statistical tools like Shannon-entropy and the  Jensen--Shannon statistical complexity \cite{enviado}.
 The authors showed that the pertinent statistical results agree with purely dynamical ones  \cite{enviado}.

\nd Our model exhibits a particularly complex sub-regime, with  superposition of chaos and complexity.
 Therein one encounters strong correlation between classical and quantum degrees of freedom \cite{KR.18}.

\subsection{Our goal}

 Statistical quantifiers often allow for interesting insights into the intricacies of purely dynamical issues
\cite{zunino}.  In such a light, the purpose of
 the present effort is to look for broader horizons in our statistical research, than those of \cite{enviado}.
This is  precisely why we appeal   to a possible q-statistics' contribution to the problem, by recourse to the q-Entropy (\ref{eq:t-entropy}) and the q-statistical complexity  \cite{qcomplexity}, that allow for considerable enlargement of our statistical arsenal.

\subsection{Methodology}

\nd Our all important PDs are extracted from times series with the BP methodology, while the time series are obtained from the  Poincare-sections arising from a non linear system of equations, that represents the extant dynamics.

\vskip 3mm

Section 2 deals with our semi-quantum system's dynamics. In
particular, Subsect. 2.1 gives results for the isolated quantum
system while  Subsect. 2.2 does so for the composite system. In
Sect. 3 we exhaustively analyze our q--information quantifiers while
Sect. 4 displays the pertinent results. Finally, some conclusions
are drawn in Sect. 5.

\section{Matter--Field Hamiltonian }

\nd Focus attention upon  the Hamiltonian \cite{KR.18}
\begin{equation}\label{Hparti}
H=\varepsilon_+(b_{+}^{\dagger }b_{+} + \frac{1}{2})+\varepsilon_-(b_{-}^{\dagger }b_{-}+ \frac{1}{2}) + (\Delta + \alpha X) \,(b_{+} b_{-}+b_{-}^{\dagger}b_{+}^{\dagger}) + \frac{\omega }{2}(P_X^{\,\,\,2}
+X^{\,2}),
\end{equation}
where $b^\dagger_\pm$, $b_{\pm}$ are boson creation and annihilation operators satisfying the standard
commutation relations ($[b_\mu,b^\dagger_{\nu}]=\delta_{\mu\nu}$, $[b_\mu,b_\nu]=[b^\dagger_{\mu},b^\dagger_{\nu}]=0$ for $\mu,\nu=\pm$), while $\varepsilon_{\pm}>0$ are the single boson energies,  and $X$, $P_X$ represent  classical coordinate and momentum quantities, with $\omega$ the associated oscillator's frequency.

\nd The quantum dynamical equations are the canonical ones \cite{K0,K1},
that is,  arbitrary operators $O$ evolve  in the Heisenberg picture  as
\begin{equation}
 i \frac{d O }{dt} = - [\: H,  O \:]\,.
\label{Eccanoncero}
\end{equation} The pertinent  evolution equation for the  mean value
$\langle O\rangle\equiv {\rm Tr}\,[\rho\,O(t)]$ becomes
\begin{equation}
i \frac{d \langle  O \rangle}{dt} = - \langle[\:  H,
  O \:]\rangle,
\label{Eccanon}
\end{equation}
with the average being  taken with respect to a proper quantum density matrix $\rho$.
Moreover,   classical variables obey the classical Hamilton's
 equations of motion
\begin{subequations}
\label{eqclasgen}
\begin{eqnarray}
\frac{dX}{dt} & = & \frac{\partial \langle  H \rangle}{\partial P_X}
\label{ds},\\
\frac{dP_X}{dt} & = & - \frac{\partial \langle  H \rangle}{\partial X}. \label{clasgenb}
\end{eqnarray}
\end{subequations}

\nd The set of equations (\ref{Eccanon}) + (\ref{eqclasgen})  is an autonomous one of coupled, first-order ordinary differential equations (ODE), that permits a dynamical description such that no quantum rule is violated. Particularly,  commutation-relations are trivially time-conserved, since
the quantum  evolution is the canonical one for our effective time-dependent Hamiltonian. Note that $X$ can be viewed as a  time-dependent parameter of our quantal system. The initial conditions are determined by a the  quantum density matrix $\rho$. Pass now to the hermitian operators
$N=b_{+}^{\dagger}b_{+}+b_{-}^{\dagger}b_{-}\,,\;\delta N=b_{+}^{\dagger}b_{+}-b_{-}^{\dagger}b_{-}\,,
O_{+}=b_{+} b_{-}+b_{-}^{\dagger}b_{+}^{\dagger} \,, \\ O_{-}=i(b_{+} b_{-}-b_{-}^{\dagger}b_{+}^{\dagger})\,,
$
and we are able to recast our Hamiltonian (\ref{Hparti})  as
\begin{equation}\label{Hparti2}
H=\varepsilon \, (N+1)+\gamma\, \delta N + (\Delta+\alpha X)\,O_{+} +\frac{\omega }{2}(P_X^{\,\,\,2}
+X^{\,2}),
\end{equation}
where $\varepsilon=(\varepsilon_++\varepsilon_-)/2>0$ and $\gamma=(\varepsilon_+-\varepsilon_-)/2$,
with $|\gamma|<\varepsilon$.
From Eqs.\ (\ref{Eccanon})--(\ref{eqclasgen}) we thus encounter a  closed system of equations for our
 set of
quantum mean values plus classical variables:
\begin{subequations}
    \label{eqquant1}
    \begin{eqnarray}
    \frac{d\langle N+1\rangle }{dt} &=&2(\Delta + \alpha X) \langle O_{-}\rangle , \label{8a} \\
    \frac{d\langle O_{-}\rangle }{dt} &=&2(\Delta + \alpha X) \, \langle N+1\rangle  +2 \varepsilon\langle
    O_{+}\rangle , \label{8b}\\
    \frac{d\langle O_{+}\rangle }{dt} &=&- 2 \varepsilon \langle
    O_{-}\rangle ,\label{8c}\\
\frac{dX}{dt} &=&\omega P_X,\label{8d} \\
\frac{dP_X}{dt} &=&-(\omega X+\alpha \langle O_{+}\rangle) ,\label{8e}
    \end{eqnarray}
\end{subequations}
where $d\langle \delta N\rangle/dt=0$.

\nd Eqs.\ (\ref{eqquant1})  are clearly  a nonlinear ODEs set. Non-linearity has been inserted  via the coupling  between the two systems, governed by the parameter $\alpha$. For $\alpha=0$
the two systems become decoupled, of course,  and the precedent  equations become, as a consequence, those for  two independent  linear systems.

\nd The expectation value $\langle O_{-}\rangle $
is regarded as a ``current'', while $\langle O_{+}\rangle $ yields the
mean value of the quantum component of the interaction
potential. Each level population
is fixed by $\langle b^\dagger_{\pm}b_{\pm}\rangle=(\langle N\rangle\pm\langle\delta N\rangle)/2$.
The full system (\ref{eqquant1})  displays moreover  the  Bloch-like  motion-invariant

 \begin{equation}\label{Inv}
I = \langle N+1 \rangle^{2} - 4|\langle b_+b_-\rangle|^2= \langle N+1 \rangle^{2}-\langle  O_{-} \rangle^{2} - \langle  O_{+} \rangle^{2},
\end{equation}
that fulfills $dI/dt=0$  in both the linear ($\alpha=0$) and nonlinear ($\alpha\neq 0$) instances, as it  is easily verified.

\nd Given that  $\langle \delta N\rangle$ is conserved, it makes  sense to work with the effective energy $E_{\rm eff}=\langle H \rangle-\gamma \, \langle \delta N \rangle -\varepsilon$ in place of
 the total energy $\langle H \rangle$. The two quantities are motion-invariants.
Employing $I$ together with $E_{\rm eff}$, we diminish the amount of freedom-degrees
of the system (\ref{eqquant1}) to just three, which enables the employment of important tools like the  {\it Poincare sections} so as to investigate the system's dynamics.

\subsection{Quantum subsystem}

\nd    For $\alpha=0$, the quantum systems  is fully described  by the quantum Hamiltonian
\begin{equation}
H_q=\varepsilon_+(b_{+}^{\dagger }b_{+} + \frac{1}{2})+\varepsilon_-(b_{-}^{\dagger }b_{-}+ \frac{1}{2}) + \Delta\,(b_{+} b_{-}+b_{-}^{\dagger}b_{+}^{\dagger})\,.
\end{equation}
The dynamics of this system is analyzed using  a method advanced  in \cite{RK.05,RK.09}, that allows for diagonalization of general quadratic forms, even if they lack positivity. The pertinent dynamics displays   {\it three}  different regimes,  according to the relation  $\Delta$ - $\varepsilon$ \cite{RK.05}.
A) One has a stable regime, for  $|\Delta|<\varepsilon$, with an evolution that  is {\it bounded and quasi-periodic}. The system can be separated into two traditional  normal modes. This regime can further  be divided into three sub-regimes according to the $H-$spectrum \cite{RK.05}. Always, discreteness and quasi-periodicity prevail  (see \cite{RK.05}). B) A dynamically unstable one,   for  $|\Delta|>\varepsilon$. The dynamics is {\it exponentially unbounded}.  The system can be split up into two normal modes. However, the creation and annihilation operators for them  are non-hermitian (see \cite{RK.05}). C) A non-separable case for $|\Delta|=\varepsilon$. Here $H$ can no longer be cast as a sum of two-independent modes \cite{RK.05}. We are here at the border between the stable and unstable regimes.

\subsection{The composite system: results}

\nd   The distinct regimes above are determined by the relation amongst $\varepsilon$, $\Delta$, and $\alpha$, no matter what the initial conditions and  $\omega$'s value may be. A) For $|\alpha|\geq\varepsilon$, the dynamics is always unbounded \cite{KR.18}.  B) For $\varepsilon > |\alpha|$, the dynamics is determined by  $\varepsilon$, $\Delta$ and $\alpha$.  $\varepsilon$ competes for significance   with the two coupling constants ($\Delta$ and $\alpha$). As $\alpha$ decreases, the system tends to  a linear scenario and  the relation between $\Delta$ and $\varepsilon$ predominates. In  \cite{KR.18} one sees illustrative Poincare sections (see Figs. 2, 3, and 6 there).  For example, if $\alpha< \varepsilon$ remains fixed but the ratio $\varepsilon/\Delta$ changes, one sees that if $\varepsilon > |\Delta|$ the dynamics is periodic and becomes quasi-periodic in the vicinity of the non-diagonalizable regime  $\varepsilon=|\Delta|$, exhibiting increasing nonlinear artifacts as this region is reached (Fig.\ 2c of  \cite{KR.18}). If $\varepsilon < |\Delta|$  un-boundedness reigns. One detects identical behavior for distinct values of $\alpha<\varepsilon$, if we keep the same ratio $\varepsilon/\Delta$. For augmenting values of $\alpha/\Delta$. Again,    evolution  from
periodic curves to rather complex, quasi-periodic ones is appreciated. Finally,  one reaches chaos.

\nd   The most remarkable  behavior is detected at the critical case  $\varepsilon \simeq|\Delta|$,
 in the  vicinity of the non-separable instance of the linear system and  at the border with the unbounded
 region. We discover  complex, quasi-periodic evolution curves. Additionally, for appropriate ``small'' values of $\alpha$ ($\alpha<\Delta$),  chaos is seen to emerge.

\section{q--Entropy and q--Statistical Complexity}
\label{sec:Information}

We are interested in physical processes described by a PD  $P = \{ p_j, j=1, \cdots, N \}$, where $N$ is  the number of available states of the physical system.
\nd     We consider the normalized q--Entropy ${\mathcal
H_{q}}$ as
\begin{equation}
{\mathcal H_{q}}[P] = {\rm S_{q}}[P]~/~{\rm S_{q}}[P_e]  \ ,
\end{equation}
where $S_q$ is given by (\ref{eq:t-entropy}) and
\begin{equation}
{\rm S_q}[P_e] = {{1-N^{1-q}} \over {q-1}} \ ,
\end{equation}
 the entropy corresponding to the uniform distribution $P_e$, for $q \in (0,1) \cup (1,\infty)$. In the Shannon case, the entropy is given by Eq. (\ref{Shannon-entropy}) ($q=1$ case) and ${\rm S_1}[P_e] = \ln N$.

As a second information measure we  will use   the  product form for
the statistical complexity  advanced  in \cite{LMC95}, ${\mathcal C}[P]~=~{\mathcal H}[P] \cdot {\mathcal Q}[P] \ $, where ${\mathcal H}[P]$ is an entropy and  ${\mathcal Q}[P]$ a distance between $P$ y $P_e$. In our case
\begin{equation}
{\mathcal C_q}[P]~=~{\mathcal H_q}[P] \cdot {\mathcal Q_q}[P] \ ,
\label{C-definition}
\end{equation}
where  ${\mathcal Q_q}[P]$ is called the q-disequilibrium,  defined  \cite{ours} via the Jensen--Tsallis divergence
 ${\mathcal J}_{{\rm S}_{q}}$ \cite{Lamberti04}
\begin{equation}
{\mathcal J}_{{\rm S}_{q}}[P,Q]~=~{{1}\over{2}}~K_{q} \left[ P,{(P + Q) \over{2}}\right]  +
                                            {{1}\over{2}}~K_{q} \left[ Q , {(P + Q) \over{2}}\right] \ ,
\label{Jensen-Tsallis}
\end{equation}
 which is the symmetric form of the
q--Kullback-Leiber relative entropy
\begin{equation}
\label{qrelative} K_{q} \left[ P ,Q \right]= \frac{1}{q-1}\; \; \; \sum_{i=1}^n p_i
\, [(\frac{p_i}{q_i})^{q-1} -1],
\end{equation}
 for $q \in (0,1) \cup (1,\infty)$. In the Shannon case, we have the  Kullback-Leiber relative entropy
\begin{equation} \label{KLsn}
K \left[ P ,Q \right]= \sum_{i=1}^n p_i
\ln\left(\frac{p_i}{q_i}\right).
\end{equation}
The square root of  ${\mathcal J}_{{\rm S}_{q}}$  is a metric \cite{Lamberti04}.
 We take
\begin{equation}
{\mathcal Q_q}[P] = {\mathcal Q_q}_0 \cdot {\mathcal J}_{{\rm S}_{q}}[P, P_e]  \ ,
\label{Q-definition}
\end{equation}
where ${\mathcal Q_q}_0$ is  a normalization constant ($0 \le
{\mathcal Q_q} \le 1$).
\begin{equation}
{\mathcal Q_q}_0 = (1-q) \cdot  \left\{ 1 - \left[{{(1+N^q)(1+N)^{(1-q)}+(N-1)}\over{2^{(2-q)}N}}
                                                \right] \right\}^{-1} \ ,
\label{q0-jensen-2}
\end{equation}
and
\begin{equation}
Q_0 = -2 \left\{ \left( {{N+1}\over{N}} \right) \ln (N+1) - \ln (2N)  +  \ln N ) \right\}^{-1} \ ,
\label{q0-jensen-1}
\end{equation} in the Jensen Shannon case. The maximum  disequilibrium
obtains  when one of the components of $P$, say $p_k$, is
unity and the remaining components vanish. The
disequilibrium ${\mathcal Q}$  reflects on the systems'
structure, becoming different from zero only  if there exist
privileged states among the available ones.

\nd   Note that    ${\mathcal
C_q}$ is not  a trivial function of the entropy. It
depends on two different probabilities distributions, namely, i)  one
associated to the system under analysis, $P$, and ii)  the uniform
distribution $P_e$. Moreover, it is  known that for a
given ${\mathcal H_q}$ value,  a range of possible SC
values can be gotten, from a minimum one ${\mathcal C_q}_{min}$  up to a maximum
value ${\mathcal C_q}_{max}$.  ${\mathcal C_q}$
provides totally original information.   A general
method  to find the bounds ${\mathcal C_q}_{min}$ and
${\mathcal C_q}_{max}$ associated  to the generalized ${\mathcal
C}={\mathcal H} \cdot {\mathcal Q}$-quantities can be encountered in Ref.
\cite{Martin2006}. Obviously,  relevant
information with regards the correlation structures among the
components of a physical system can be obtained from  the
statistical complexity quantifier.  Next, we  numerically analyze the system's dynamics using the two q-quantifiers above.
\vskip 5mm

\section{Present results}

\nd
We employ initial conditions consistent with a proper  density operator. Thus,  the uncertainty relationships of the quantum system are verified at all times. Precisely, the accuracy of our treatment  was checked out  by verifying   the time-constancy of  $E_{\rm eff}$ and $I$ (our dynamical invariants) up to  a  $10^{-10}$ precision.

\nd
Time series (TS) to build up the  PDs $P$
 are found using the systems'  Poincare sections (PS). Another procedure is to find the PDs using phase space's curves, what we also did.  Of course,  PS's are preferable representatives of phase space than  curves in it. Our present numerical results confirm this desirability.

\nd    Our PD's are extracted from the TS using the Bandt-Pompe technique (see the Appendix). The succession of PS's employed  in our computations  are gotten  via crossings with a plane, i.e.,    solutions of (\ref{eqquant1}) with the  $X(t)=0$ plane  for identical values of the invariants $E_{\rm eff}$ and $I$. We also change    $\varepsilon/\Delta$ and maintains constant  both  $\alpha/\Delta$ and  $\omega/\Delta$ in the PS's succession.

\nd    For each PS linked  to a certain $\varepsilon/\Delta >1$ we work with $21$ curves, drawn by changing the initial conditions $\langle O_{-} \rangle_0$ and $P_0$ (keeping compatibility with our  values for $E_{\rm eff}$ and $I$). In the unbounded zone ($\varepsilon/\Delta =1$) we require $10000$ curves.  $X_0$, $\langle N \rangle_0$, and $\langle O_{+} \rangle_0$ are maintained constant. Further, for each PS our TS is the one  associated  to
 time-dependent values of different quantities like $\langle N+\rangle$, $\langle  O_{-}\rangle$, $\langle  O_{+}\rangle$, etc. The graphs  depicted here are linked  to the  $\langle  O_{+}\rangle$-case.
One finds similar results for any of these quantities. We selected, per PS,  $10000$ crossing-points with the plane
 $X(t)=0$.

 \nd A consistent Shannon (+ Jensen-Shannon) statistical description of our model, that agrees with the purely dynamic one, has been presented in \cite{enviado}.
 The results can be observed, together with those corresponding to different values of $ q $ in Figs. 1--4.


\nd     Fig. 1 displays ${\mathcal H_q}$ vs. $\varepsilon/\Delta$ for different q-values, including $q=1$.
 In all cases, for decreasing  $\varepsilon/\Delta$ one sees that  ${\mathcal H_q}$ grows (with slight oscillations), from the quasi-periodic zone ($\varepsilon/\Delta> 1.2$) towards $\varepsilon/\Delta=1$, till becoming maximal at $\varepsilon/\Delta\simeq 1.05$. The dynamics teaches us that  chaoticity suddenly  emerges therein  \cite{KR.18}. Afterwards,  in all cases, ${\mathcal H_q}$ suddenly drops in the unbounded dynamics' zone  ($\varepsilon/\Delta\simeq1$) till reaching an absolute minimum at  ($\varepsilon/\Delta=1$). For ($\varepsilon/\Delta < 1$), ${\mathcal H_q}$ is close to a minimum, almost null value. One should expect that  ${\mathcal H_q}$ be smaller in this region than in the quasi-periodic (or even the non periodic) zone.
 The most noticeable  ${\mathcal H_q}$-variations emerge in the region lying between  $\varepsilon/\Delta\simeq 1.2$ and   $\varepsilon/\Delta\simeq 1.05$, associated to the entropic maximum. Dynamically, this region is linked to a region in which non-linearity becomes of a more involved nature. This tales place  as we attain $\varepsilon/\Delta=1.05$, near  $\varepsilon/\Delta=1$, value that signals the quantum unstable scenario. Remind that here we can not   find  separability  into quantum normal modes.

\nd    Fig. 2 displays the q--statistical complexity (SC) vs. $\varepsilon/\Delta$ for a smaller q-range.
Roughly,  ${\mathcal C_q}$ behaves like ${\mathcal H_q}$ for all $q$. Notice that if $\varepsilon/\Delta$ decreases, SC grows  till  $\varepsilon/\Delta\simeq 1.2$. Onwards, it strongly oscillates till $\varepsilon/\Delta\simeq 1.08$, attaining an absolute maximum. From this point onwards, ${\mathcal C_q}$ suddenly diminishes, reaching an absolute minimum in the unbounded zone.

\nd   Even if the minima are reached at the same $\varepsilon/\Delta$-value, the maxima of  ${\mathcal H_q}$  and ${\mathcal C_q}$ are not attained in the same manner. The SC reaches its maximum  \textbf{sooner} than the entropy in  the process of approaching the unstable, quantal point. Even if the concomitant  $\varepsilon/\Delta$-values   do not differ too much among themselves, they are not identical.

\nd We conclude  that the descriptions via  ${\mathcal H_q}$ and ${\mathcal C_q}$ can be regarded as reconfirming   the $q=1$-one obtained in \cite{enviado}.

\nd Fig. 3 depicts ${\mathcal H_q}$  vs. $\varepsilon/\Delta$
in the  considerable ${\mathcal H_q}$--validity range $q$ $\in$ $(0, 3.5)$. In this range, the entropy maximum is located  \textbf{in the same site } $\varepsilon/\Delta\simeq 1.05$, as in the Shannon case.
Instead, at $q=3.5$, the entropy no longer distinguishes between the dynamic-transition zone and the quasi-periodic one. This fact sets an upper limit to $q$. Fig. 3 is an illustration. For $q<1$, contrarily, these two zones are better distinguished. Also, we find there a stronger similitude between the curves for ${\mathcal H_q}$ and  ${\mathcal C_q}$. The latter loses then significance.

Questions about the validity range (VR) for ${\mathcal C_q}$ are answered by stating that its VR is much smaller that for the entropy. Now we have  $q$ $\in$ $[0.8, 1.6]$. For $q > 1.6$ the q-complexity absolute maximum is located in the quasi--periodic zone, not in the transition one. This result is not consistent with the dynamic results. This places an upper limit of $q=1.6$. As for a lower bound, we find  $q=0.8$. This is because, for $q<0.8$, the $\varepsilon/\Delta= 1.05-$value  at which the q-complexity is maximal  coincides with that of the entropy. We can the say that the q-complexity loses relevance. The location of the q-complexity maximum changes for $[0.8, 1.6]$-q-range. The changes are not large. For $q$ $\in$ $[0.8, 1.2)$ the maximum is attained at $\varepsilon/\Delta= 1.08$, as in the $q=1$ case. When $q $ grows, the location grows as well, reaching  $\varepsilon/\Delta= 1.1$ for $q=1.6$. The   optimal $\varepsilon/\Delta$-value for the ${\mathcal C_q}$-maximum  cannot be obtained with our methodology. Maybe another complexity functional might be needed.

\nd Fig. 4 depicts q-complexity curves for different $q$-values
in the VR, $[0.8, 1.6]$.

\section{Conclusions}

\nd   By recourse to Tsallis' statistical tools we studied  a non lineal Hamiltonian that describes the
interaction of a quantum--matter system with a classical field. The field is represented by a single-mode
electromagnetic one. The quantum system is a bosonic one that admits of both unbounded and quasi-periodic regimes. These
two regimes  are separated by an unstable third one \cite{RK.05}. The composite system is of
interest in quantum optics and in condensed matter \cite{Milonni,Sa.91,K0,K1}.

\nd   The dynamics of the composite system is governed by a non-linear system of ordinary differential equations
 (ODE), given by  (\ref{eqquant1}). This ODE displays periodic, quasi-periodic, unbounded, chaotic, and non-linear sub-dynamics, depending on the $H$-parameters' values. An interesting feature  is that both the complex non-linear and the chaotic sub-dynamics are found lie (in the parameters' space) in the vicinity of the unstable isolate quantum regime. Although the presence of the classical system is what enables the existence of non-linearity and chaos, one can reasonably deduce from this feature  that important model's properties emerge from the quantum system.

\nd   Our statistical tools are the q--entropy ${\mathcal H_q}$ and the q--statistical complexity ${\mathcal C_q}$, evaluated via the Bandt-Pompe symbolic analysis from time-series (TS).  A specials case (q=1) is that of the  Shannon entropy and Jensen-Shannon's complexity. In turn, the TS were obtained from Poincare sections (PS) derived via our ODE system.
 We get the PS through intersections of the ODE's solutions of (\ref{eqquant1}) with the  $X(t)=0$ plane, keeping constant the invariants $E_{\rm eff}$ and $I$. In our graphs we also keep constant i) the values of  $\alpha/\Delta$ and  $\omega/\Delta$ and ii) the initial conditions  $X_0$, $\langle N \rangle_0$ and $\langle O_{+} \rangle_0$ (for all the PS-succession). One varies   $\varepsilon/\Delta$.

\nd   As a first conclusion we have verified the sturdy nature of our results. The q-description (within a reasonable q-range), as seen in Figs. 1 - 2, is coherent with the Shannon's one. Both Shannon's entropy and ${\mathcal H_q}$ (for $0 \le q \le 3.5)$, reach an absolute maximum at the \textbf{same value} of $\varepsilon/\Delta= 1.05$.

\nd   As a second result we have found that our description's validity-range is determined by  ${\mathcal C_q}$. This range is $q$ $\in$ $[0.8, 1.6]$ (Fig. 4), more restricted than that of the q-entropic range mentioned above
 (Fig. 3).

\nd   Lastly, the ${\mathcal C_q}$-maximum's position varies between $\varepsilon/\Delta= 0.8$ and $\varepsilon/\Delta= 1.1$. The optimal ${\mathcal C_q}$-maximum's position-value cannot be ascertained by recourse to
the present information-tools. Maybe still more general entropic functionals, maybe of not trace-form, could become useful.

\textbf{Acknowledgments}. AK acknowledges support from CIC of Argentina. AP acknowledges support from CONICET of Argentina.

\appendix
\subsection{Appendix. PD Based on Bandt and Pompe's Methodology}
\label{sec:PDF-Bandt-Pompe}   To use the Bandt and Pompe
\cite{Bandt2002} methodology for evaluating the probability
distribution $P$ associated with the time series (dynamical system),
one starts by considering partitions of the pertinent
$D$-dimensional space that will hopefully ``reveal" relevant details
of the ordinal structure of a given one-dimensional time series
$\mathcal S(t) = \{ x_t; t= 1, \cdots, M\}$, with embedding
dimension $D > 1$ and time delay $\tau$.
We will take here $\tau = 1$ as the time delay, a parameter of the approach
\cite{Bandt2002}. We are  interested in ``ordinal patterns", of order
$D$ \cite{Bandt2002,Rosso2012}, generated by
\begin{equation}
(s)~\mapsto~
\left(~x_{s-(D-1)},~x_{s-(D-2)},~\cdots,~x_{s-1},~x_{s}~\right),
 \label{vectores11}
\end{equation}
which assigns to each time
the $D$-dimensional vector of values at times $s,
s-1,\cdots,s-(D-1)$. Clearly, the greater the $D-$value, the more
information on the past  is incorporated into our vectors. By
``ordinal pattern" related to the time $(s)$, we mean the
permutation $\pi=(r_0,r_1, \cdots,r_{D-1})$ of $[0,1,\cdots,D-1]$
defined by
\begin{equation}
x_{s-r_{D-1}}~\le~x_{s-r_{D-2}}~\le~\cdots~\le~x_{s-r_{1}}~\le~x_{s-r_0}.
\label{permuta11}
\end{equation}
In this way the vector defined by Eq.~(\ref{vectores11}) is
converted into a unique symbol $\hat{x}_i$. Thus, a permutation
probability distribution $P_x=\{p(\hat{x}_i), i=1,\dots, D!\}$ is
obtained from the time series $x_i$. The probability distribution
$P$ is obtained once we fix the embedding dimension $D$ and the time
delay $\tau$. The former parameter plays an important role for the
evaluation of the appropriate probability distribution, since $D$
determines the number of accessible states, $D!$, and tells us about
the necessary length $M$ of the time series needed in order to work
with a reliable statistics. The whole enterprise works for  $D! \ll M$. In particular, Bandt and
Pompe~\cite{Bandt2002} suggest for practical purposes to work with
$3 \le D \le 7$. For more details see \cite{Rosso2012}. We have considered in this work $D = 6$,
a reasonable value given in the literature for series of length $M = 10000$.
We have checked the results taking $D = 5$, obtaining similar descriptions for the information measures considered.

\newpage
\begin{figure}
    \vspace*{-1.cm}
    \centerline{\hspace*{4.cm}\scalebox{.50}{\includegraphics{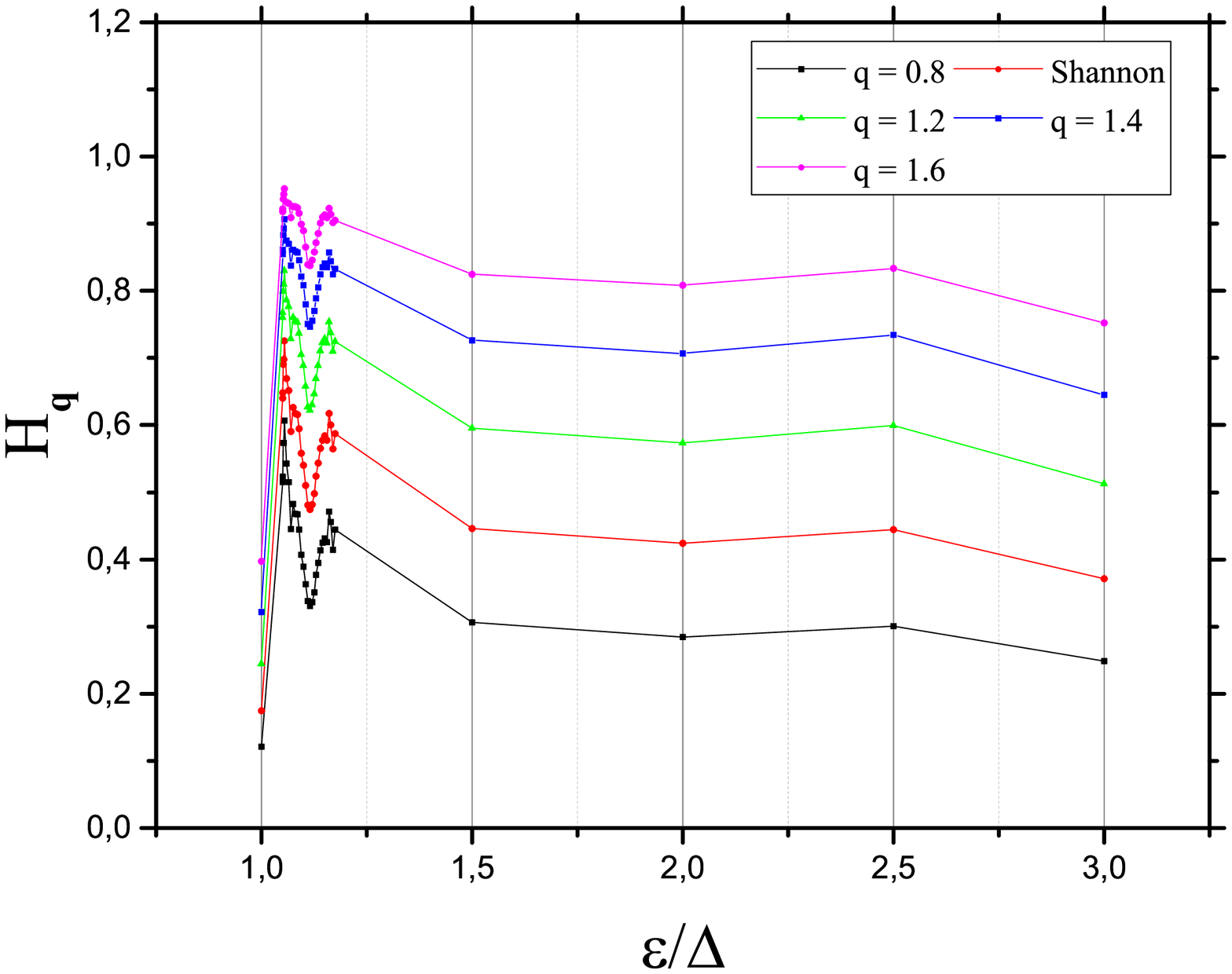}}}
    \vspace*{-3.cm}
    \caption{Entropy ${\mathcal H_q}$ vs. $\varepsilon/\Delta$  for different  $q$--values, including the Shannon case.  ${\mathcal H_q}$ is calculated with PDFs extracted from Poincare sections for the $X=0$ plane, corresponding to  $E_{\rm eff}=4.8$ and $I=4$, with $X_0=1$, $\langle N \rangle_0=1$ and $\langle O_{+} \rangle_0=0$. We set $\omega/\Delta=1$ and $\alpha/\Delta=0.015$ while  the ratios  $\varepsilon/\Delta$ change.  In all the curves, if   $\varepsilon/\Delta$ decreases,  ${\mathcal H_q}$ grows (with oscillations) from the quasi-periodic zone ($\varepsilon/\Delta> 1.2$) towards $\varepsilon/\Delta=1$. It becomes maximal at $\varepsilon/\Delta\simeq 1.05$. Chaos suddenly  emerges therein. Afterwards,  ${\mathcal H}$ suddenly drops in the unbounded dynamics' zone  ($\varepsilon/\Delta\simeq1$) till reaching an absolute minimum at  ($\varepsilon/\Delta=1$).}\label{Figs1}
\end{figure}

\begin{figure}
        \vspace*{0.cm}
    \centerline{\hspace*{4.cm}\scalebox{.50}{\includegraphics{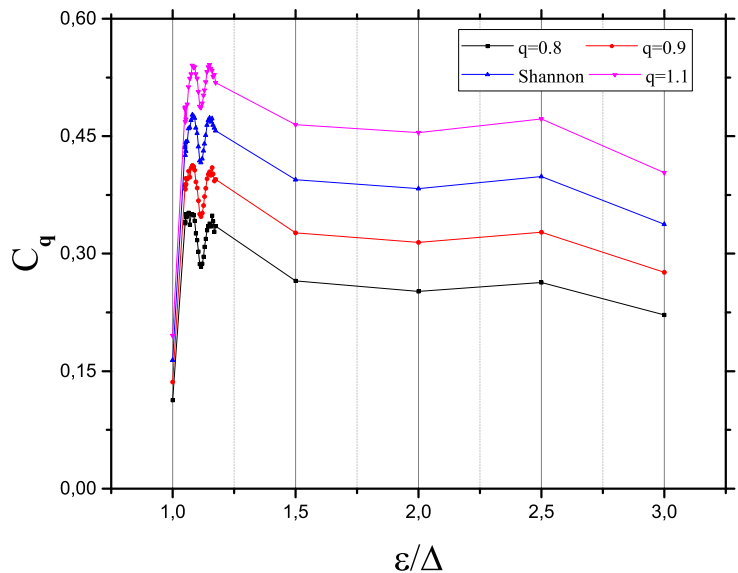}}}
    \vspace*{-3.cm}
    \caption{Statistical Complexity ${\mathcal C_q}$ vs. $\varepsilon/\Delta$, calculated as in Fig. 1. Roughly,  ${\mathcal C_q}$ behaves like ${\mathcal H_q}$. For all $q$, see that if $\varepsilon/\Delta$ decreases, ${\mathcal C_q}$ grows  till  $\varepsilon/\Delta\simeq 1.2$. Onwards, it strongly oscillates till $\varepsilon/\Delta\simeq 1.08$, reaching an absolute maximum. Herefrom, ${\mathcal C_q}$ suddenly diminishes, reaching an absolute minimum in the unbounded zone.}\label{Figs2}
\end{figure}

\newpage

\begin{figure}
    \vspace*{-1.cm}
    \centerline{\hspace*{4.cm}\scalebox{.50}{\includegraphics{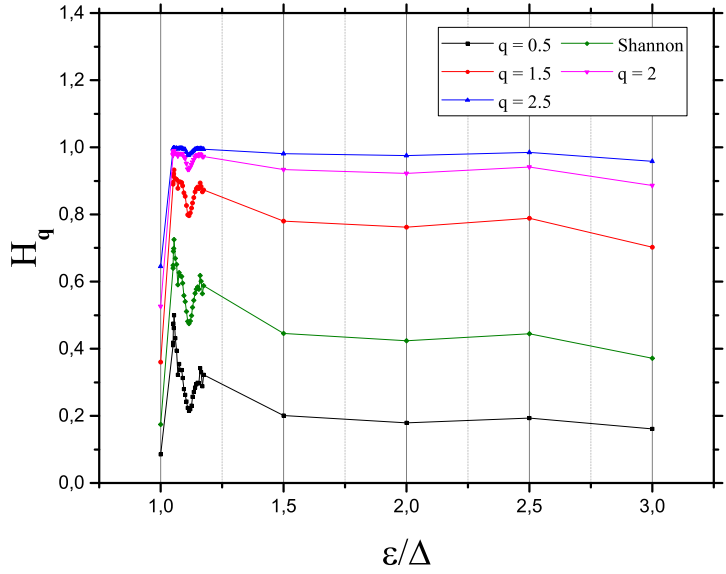}}}
    \vspace*{-3.cm}
    \caption{${\mathcal H_q}$  vs. $\varepsilon/\Delta$, calculated as in Fig. 1, but in the ${\mathcal H_q}$--validity range $q$ $\in$ $(0, 3.5)$. In this range, the entropy maximum is located  \textbf{at the same value } of  $\varepsilon/\Delta\simeq 1.05$, as in the Shannon case.}\label{Figs3}
\end{figure}

\begin{figure}
        \vspace*{-2.cm}
    \centerline{\hspace*{4.cm}\scalebox{.50}{\includegraphics{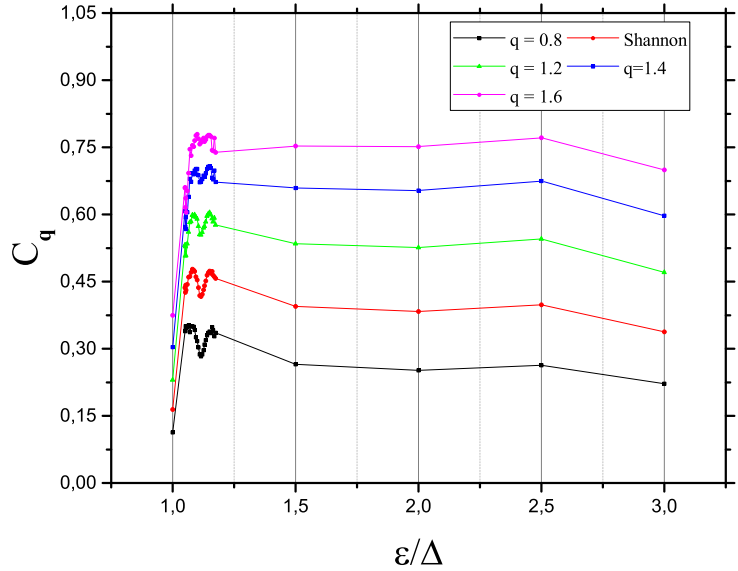}}}
    \vspace*{-3.cm}
    \caption{We plot ${\mathcal C_q}$ vs. $\varepsilon/\Delta$, as in Fig. 2, but for different $q$-values in the ${\mathcal C_q}$--validity range, $[0.8, 1.6]$. The location of the q-complexity maximum changes in this range, but the changes are not significant.}\label{Figs4}
\end{figure}

\end{document}